\documentclass[AMA,STIX1COL]{WileyNJD-v2}

\articletype{Research Article}%

\received{26 April 2016}
\revised{6 June 2016}
\accepted{6 June 2016}

\usepackage{units}
\usepackage{caption}
\usepackage[font=normalsize, labelfont=normalsize, textfont=normalsize]{subcaption}
\usepackage{tikz}
\usetikzlibrary{calc}
\usetikzlibrary{decorations.markings}
\usepackage{pgf}
\usepackage{pgfplots} 
\usepackage{pgfplotstable}
\usepackage{todonotes}
\usepackage{amsmath}
\usepackage{amsfonts}
\graphicspath{{Figures/}}

\newcommand*\diff{\mathop{}\!\mathrm{d}}

\newcommand{\executeiffilenewer}[3]{%
	\ifnum\pdfstrcmp{\pdffilemoddate{#1}}%
	{\pdffilemoddate{#2}}>0%
	{\immediate\write18{#3}}\fi%
}

\newenvironment{rcases}
{\left.\begin{aligned}}
	{\end{aligned}\right\rbrace}

\definecolor{rwth}   {RGB}{  0  84 159}
\definecolor{bordeaux}   {RGB}{161  16  53}
\definecolor{rwth-75}{RGB}{ 64 127 183}
\definecolor{rwth-50}{RGB}{142 186 229}
\definecolor{rwth-25}{RGB}{199 221 242}
\definecolor{rwth-10}{RGB}{232 241 250}
\definecolor{petrol}   {RGB}{  0  97 101}
\definecolor{violett}   {RGB}{ 97  33  88}
\definecolor{grun}   {RGB}{ 87 171  39}
\definecolor{rot}   {RGB}{204   7  30}

\definecolor{orangenew}   {RGB}{218   124  48}
\definecolor{lilanew}   {RGB}{107   76  154}
\definecolor{graynew}   {RGB}{83   81  84}
\definecolor{olivenew}   {RGB}{148   139  61}

\colorlet{colorI}{rot}
\colorlet{colorII}{rwth}
\colorlet{colorIII}{grun}
\colorlet{colorIV}{orangenew}
\colorlet{colorV}{violett}
\colorlet{colorVI}{olivenew}
\colorlet{colorVII}{graynew}

\begin{document}
	
	\title{Simulating dense granular flow using the $\mu$($I$)-rheology within a space-time framework\protect}
	
	\author{Linda Gesenhues*}
	
	\author{Marek Behr}

	\authormark{Gesenhues and Behr}

	\address{\orgdiv{Chair for Computational Analysis of Technical Systems (CATS)},\linebreak \orgname{RWTH Aachen University},\linebreak \orgaddress{\state{Aachen}, \country{Germany}}}

	\corres{*Linda Gesenhues, Chair for Computational Analysis of Technical Systems (CATS), \linebreak RWTH Aachen University, \linebreak 52056 Aachen, Germany.  \linebreak \email{gesenhues@cats.rwth-aachen.de}}

	\abstract[Summary]{
		A space-time framework is applied to simulate dense granular flow. Two different numerical experiments are performed: a column collapse and a dam break on an inclined plane.  The experiments are modeled as two-phase flows. The dense granular material is represented by a constitutive model, the $\mu$($I$)-rheology, that is based on the Coulomb's friction law, such that the normal stress applied by the pressure is related to the tangential stress. The model represents a complex shear thinning viscoplastic material behavior. The interface between the dense granular material and the surrounding light fluid is captured with a level set function. Due to discontinuities close to the the interface, the mesh requires a sufficient resolution. The space-time approach allows unstructured meshes in time and, therefore a well refined mesh in the temporal direction around the interface. In this study, results and performance of a flat and a simplex space time discretization are verified and analyzed.
	}
	

	\maketitle

	\section{Introduction}\label{sec:Introduction}
	Dense granular flows appear in various occasions in industry and nature. Typical industry applications are storage and processing of grains, like rice and lentils, or tablet handling in the pharmaceutical sector. In nature, dense granular flow occurs in debris flows, turbidity currents, lava, and mud flows. 
	
	Computational fluid dynamics can help to understand and predict events that involve dense granular flow. Hereby, the material properties of the dense granular material need to be characterized. This can be done by a continuous material model. The $\mu$($I$)-rheology is a constitutive model that describes the viscosity of dense granular material. It is an empirical model based on phenomena observed in experiments \cite{Midi2004,Pouliquen2002,Jop2005}. This model represents a complex shear thinning viscoplastic behavior that is based on the Coulomb's friction law, such that the normal stress applied by the pressure is related to the tangential stress. However, the high non-linearity of the viscosity model and its broad range of viscosity values make the numerical simulation challenging. In addition, simulation of dense granular flows is often coupled with a two-phase or free surface problem. To capture viscosity gradients and the flow front accurately, a well refined finite element mesh is necessary. Yet, a high resolution leads also to high computational cost. Adaptive mesh refinement can provide a sufficient resolution in the areas of interest paired with good efficiency. In this way, in the region at interest, a very fine mesh is applied, whereas the mesh elsewhere remains coarse. Mesh refinement includes the refinement of the spatial mesh, or for space-time meshes, the refinement in temporal direction. The latter applies smaller effective time steps in the refined areas, and bigger time steps elsewhere.
	
	Widely used benchmarks to verify and develop methods related to dense granular flow are a column collapse or a dam break.
	Lagr\'{e}e \textit{et al.}~\cite{Lagree2011} perform a discrete element and a continuous simulation using the $\mu$($I$)-rheology of a column collapse. Further studies investigated the 2D and 3D column collapse  also taking in consideration the sensitivity  of the empirical and regularization parameters. \cite{Gesenhues2019,Gesenhues2017,Riber2016,Valette2019,Lin2020,Rauter2020,Ke2020}
	
	A column collapse or dam break simulation needs to deal with the interface or free surface of the dense granular fluid. Here, a Lagrangian, Eulerian, or an arbitrary Lagrangian-Eulerian framework can be used. This study uses an Eulerian approach where the level set formulation tracks the interface between the dense granular fluid and the light surrounding fluid. For a precise position of the interface, a well refined mesh and a small time step are crucial. In this study, we use an unstructured space-time approach to simulate a 2D column collapse incorporating the $\mu$($I$)-rheology as well as a 2D dam break on an inclined plane.

	The space-time approach is an alternative to the classical finite difference temporal discretization schemes like the $\theta$-methods or Runge-Kutta methods. Here, the temporal domain is, like the spatial domain, approximated with a finite element scheme. A space-time framework can be advantageous as it is third-order accurate even for a simple linear-in-time interpolation. Moreover, a space-time discretization using simplex elements allows unstructured meshes in time. Therefore, it can be used for a better resolution of discontinuities at the interface and for topology changes.
	Previous studies showed how simplex elements can be built and how space-time can be used to simulate topology changes. Examples are a rising bubble and a cavity filling. \cite{Erickson2005, Behr2008, Wang2013, Lehrenfeld2015, Karyofylli2018, Karyofylli2018a, VonDanwitz2019, Karyofylli2019}
	
	This study shows the results of a column collapse and an inclined dam break using a flat and simplex space-time approach. The column collapse is simulated for two different aspect ratios. The flow fronts are verified and performance is analyzed. The inclined dam break simulates a dense granular flow down an inclined plane against a rigid obstacle and compares the impact time and impact force on the obstacle.
	
	The paper is organized as follows: In Sec.~\ref{sec:Methods}, the governing and constitutive equations are presented. Section~\ref{sec:FEM} states the finite element discretization and more information on the structure of space-time elements.  Section~\ref{sec:ColumnCollapse} shows the results of the column collapse. In Sec.~\ref{sec:Bulldoze}, simulation results of the inclined dam break are presented. Section~\ref{sec:Conclusion} holds a summary and a conclusion.
	
	%
	%
	%
	%

	\section{Methods}\label{sec:Methods}
	\subsection{Governing equations} \label{sec:GoverningEquations}
	On a spatial domain $ \Omega \subset R^3$, with a piecewise regular surface and time interval $[0,t_f]$, an incompressible and viscous fluid is governed by the momentum and continuity equations:
	\begin{align}
	\rho \left( \frac{\partial \mathbf{u}}{\partial t} +    \mathbf{u} \cdot \nabla  \mathbf{u}\right) - \nabla \cdot \boldsymbol{\sigma} = \mathbf{f} \quad & \mbox{on } \Omega \times [0,t_f] \label{eq:momentum_NS} \, ,\\  
	\mbox{\boldmath $\nabla$} \cdot \mathbf{u} = 0 \quad & \mbox{on } \Omega \times [0,t_f] \label{eq:continuity_NS} \, ,
	\end{align}
	where $\mathbf{u}$ represents the velocity field, $\rho$ is the fluid density, $ \boldsymbol{\sigma} $ holds the stress tensor, and $\mathbf{f}$ denotes the body force.
	The essential and natural boundary conditions are described as:
	\begin{align}
	\mathbf{u}=\mathbf{g} \quad & \mbox{on }\, \Gamma_{\mathbf{g}}\, , \\
	\mathbf{n} \cdot \boldsymbol{\sigma} =  \mathbf{h} \quad & \mbox{on }\, \Gamma_{\mathbf{h}} \, ,
	\end{align}
	where $\Gamma_{\mathbf{g}}$ and $\Gamma_{\mathbf{h}}$ are complementary subsets of the boundary $\Gamma$. The Dirichlet boundary condition is given by $\mathbf{g}$, and $\mathbf{h}$ represents the Neumann boundary condition. The vector $\mathbf{n}$ denotes the unit outward normal vector of $\Gamma$. The initial condition is:
	\begin{equation}
	\mathbf{u}\left(\mathbf{x},0\right)=\mathbf{u}_{0}\, ,
	\end{equation}
	where $\mathbf{u}_{0}$ is divergence-free.
	The stress tensor $ \boldsymbol{\sigma} $ can be described by:
	\begin{equation}\label{eq:constitutive}
	\boldsymbol{\sigma}(p,\mathbf{u}) = -p\mathbf{I} + \mathbf{T}\, ,
	\end{equation}
	where $\mathbf{I}$ is the identity tensor and $p$ denotes the pressure. The deviatoric part of the stress tensor $\mathbf{T}$ for a generalized Newtonian fluid is:
	\begin{equation}\label{eq:newtonian}
	\mathbf{T}=2 \eta \boldsymbol{\varepsilon}(\mathbf{u})\, .
	\end{equation}
	Here, $\eta$ describes the dynamic viscosity and $\boldsymbol{\varepsilon}$ holds the strain rate tensor, which is defined as:
	\begin{equation}
	\boldsymbol{\varepsilon} =  \frac{1}{2}(\nabla \mathbf{u} + \left(\nabla \mathbf{u}\right)^T) \, .
	\end{equation}
	Consequently, Eq.~\eqref{eq:momentum_NS} can be restated as:
	\begin{align}
	\rho \left( \frac{\partial \mathbf{u}}{\partial t} + \mathbf{u} \cdot \nabla  \mathbf{u}\right) - \nabla \cdot \left(2\eta \boldsymbol{\varepsilon}\left(\mathbf{u}\right)\right)+ \nabla p= \mathbf{f} \quad  \mbox{on } \Omega \times[0,t_f]\, .  
	\end{align}
	For a Newtonian fluid, the dynamic viscosity $\eta$ is constant. Thus the relationship between the stress tensor $\boldsymbol{\sigma}$ and the strain rate tensor $\boldsymbol{\varepsilon}$ is linear. However the viscosity of a non-Newtonian fluid may depend on other flow parameters such as shear rate or pressure. So, Eq.~\ref{eq:newtonian} may be written as:
	\begin{equation}\label{eq:VOF}
	\mathbf{T}=2 \eta(\dot\gamma,p) \boldsymbol{\varepsilon}(\mathbf{u})   \, ,
	\end{equation}
	where the shear rate $\dot\gamma$ is calculated from the second invariant of the strain rate tensor:
	\begin{equation}
	\dot\gamma = \left(\frac{1}{2}\boldsymbol{\varepsilon}\colon\boldsymbol{\varepsilon}\right)^{\frac{1}{2}} \, .
	\end{equation}
	\subsection{$\mu$($I$)-Rheology}
	The $\mu$($I$)-rheology is based on the Coulomb friction, which assumes that the tangential stress $\tau$ is proportional to the normal stress. In the case of the $\mu$($I$)-rheology, the normal stress is applied by the pressure. So that $\tau=\mu p$, where $\mu$ is a friction coefficient. However, the $\mu$($I$)-rheology relates the friction parameter to the flow by the inertial number $I$. This number is non-dimensional and represents the local state of the granular packing:
	\begin{equation}\label{eq:inertialnumber}
	I=\frac{ \dot\gamma d}{\sqrt{\left(p/\rho\right)}}  \, ,
	\end{equation}
	where $d$ is the grain diameter and $\rho$ denotes the density. The inertial number $I$ can be interpreted as a ratio of  the microscopic timescale $\sqrt{\left(d^{2}\rho/p\right)}$ and the macroscopic deformation timescale ($1/\dot\gamma$). Thus, the dynamic viscosity $\eta$ is given by:
	\begin{equation}\label{eq:MuViscosity}
	\eta= \frac{\mu(I)p}{\dot\gamma} \, .
	\end{equation}
	Jop \textit{et al.} \cite{Jop2005} propose a law for the friction coefficient based on experiments executed by Pouliquen and Forterre \cite{Pouliquen2002,Pouliquen1999}. The friction coefficient $\mu(I)$ starts from a critical friction value $\mu_{s}$ and converges towards a limiting friction value $\mu_{2}$, so that:
	\begin{equation}\label{eq:mu(I)}
	\mu\left(I\right)_{\mathit{Jop}}= \mu_{s}+\frac{\mu_{2}-\mu_{s}}{I_{0}/I +1} \, .
	\end{equation}
	Here, $\mu_{s}$, $\mu_{2}$, and $I_{0}$ are empirical quantities determined by experiments.
	
	The complexity of the $\mu$($I$)-rheology law requires several regularization steps. The $\mu$($I$)-rheology is ill-posed and suffers from the unlimited growth of small perturbations\cite{Barker2015}. Barker and Gray \cite{Barker2017} suggested a partial regularization of the $\mu$($I$)-rheology:
	\begin{align}\label{eq:barker2}
	\mu\left(I\right)=\begin{cases}
	\sqrt{\frac{\alpha}{\ln\left(\frac{A\_}{I}\right)}} \quad& I \leqslant I_{1}^{N}  \, , \\
	\mu\left(I\right)_{\mathit{Jop}} \quad & I > I^{N}_{1}  \, ,
	\end{cases}
	\end{align}
	where $I^{N}_{1}$ and $I^{N}_{2}$ denote the neutral inertial numbers and $\alpha$  represents a fitting parameter $\alpha\lesssim 2$. Note that the limiting neutral inertial numbers $I^{N}_{1}$ and $I^{N}_{2}$ depend exclusively on the rheological parameters of the $\mu$($I$)-rheology and not on the flow conditions.
	In this study, $\alpha=1.5$ is used. The coefficient $A\_$ is given as:
	\begin{equation}
	A\_=I_{1}^{N}\exp\left(\frac{\alpha}{\mu_{\mathit{Jop}}(I_{1}^{N})^{2}}\right) \, .
	\end{equation}
	
	Besides the regularization for low inertial numbers, further regularizations in the form of viscosity thresholds are necessary to enhance convergence during the simulation. 
	Considering that the denominator of each term of the viscosity function cannot be equal to zero and reviewing Eqs.~\eqref{eq:inertialnumber}, \eqref{eq:MuViscosity}, and \eqref{eq:mu(I)}, it follows that if the pressure $p$, shear rate $\dot\gamma$, or the inertial number $I$ are zero, the function does not return valid results. For such cases, several regularization strategies can be applied \cite{Frigaard2005,Chauchat2014,Franci2018}.
	In this study, a simple regularization parameter $\lambda=10^{-5}$ is added, consisting of a small regularization parameter, to the denominators.  
	Moreover, the viscosity is bounded:
	\begin{equation}\label{eq:viscositylimit}
	\eta=min\{max\{\eta_{min}, \,\eta\left(\dot\gamma, \,p   \right)\}\,,\eta_{max}\} \,,
	\end{equation} 
	with $\eta_{min}$ and $\eta_{max}$ as input parameters.
	The choice of the minimum and maximum viscosity is a compromise between convergence and accuracy \cite{Gesenhues2017}.
	
	\subsection{Level Set Method}
	
	The level set is a method that captures the interface between two fluids with a smooth level set function that copes with the discontinuous viscosity and density. \cite{Chang1996,Sussman1999}. The fluid velocity drives the level set marker $\phi$ according to the following transport equation:
	\begin{equation}\label{eq:levelset}
	\frac{\partial \phi}{\partial t} + \mathbf{u} \cdot \mbox{$\nabla$} \phi = 0 \quad \mbox{on } \Omega \times [0,t_f] \, .
	\end{equation}
	The marker $\phi$ defines the state of the fluid, as for positive $\phi$, the heavy fluid properties are assumed and for negative $\phi$, the light fluid is considered.
	\begin{align}
	\rho = \begin{cases}
	\rho_H \quad \text{for} \; \phi\left(\mathbf{x},t\right)<0\,, \\
	\rho_L \quad  \text{for}  \; \phi\left(\mathbf{x},t\right)>0\,,
	\end{cases}
	\eta = \begin{cases}
	\eta_H \quad \text{for} \; \phi\left(\mathbf{x},t\right)<0\,, \\
	\eta_L \quad  \text{for}  \; \phi\left(\mathbf{x},t\right)>0\,.
	\end{cases}
	\end{align}
	The interface is defined at $\phi\left(\mathbf{x},t\right)=0$. Ideally, the level set function stores the shortest distance from a point $\mathbf{x}$ for each time instance to the interface, such that:
	\begin{equation}
	\phi\left(\mathbf{x},t\right)=\pm \min_{\mathbf{x^{*}\in\Gamma^{int}_{t}}}\lVert\mathbf{x}-\mathbf{x^{*}}\rVert \, , \quad \forall\mathbf{x}\in \Omega\,.
	\end{equation}
	where $\mathbf{x^{*}}$ belongs to the interface.
	\section{Finite element formulation and numerical methods for a space-time framework} \label{sec:FEM}

	\subsection{Finite element formulation of the Navier-Stokes equations}
\begin{figure}[h!]
		\centerline{
			\includegraphics[width=0.4\textwidth]{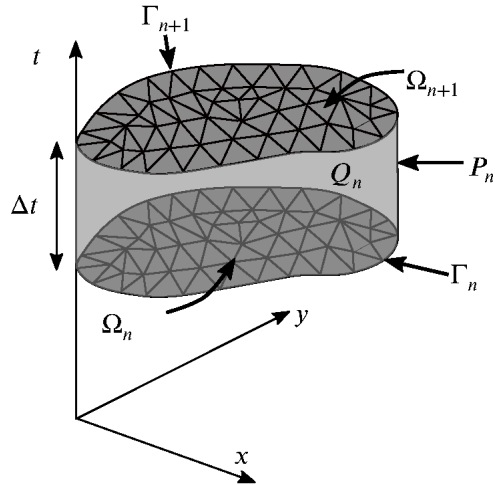}}
		\caption{Space-time slab for a 2D spatial domain.}
		\label{fig:spaceslab}		
	\end{figure}

	The space-time approach uses a finite element formulation for the space-time domain $Q$. The time interpolation is discontinuous, so that two values of the unknowns are saved for each node. The time interval $t\in \left[0,T\right]$ is split in subintervals $I_{n}=\left(t_{n},t_{n+1}\right)$ such that $t_{n+1}=t_{n}+\Delta t$. The spatial domain $\Omega_{n}=\Omega\left(t_{n}\right)$ has a boundary $\Gamma_{n}=\Gamma\left(t_{n}\right)$. The time-slab $Q_{n}$ is defined by the domain that is embedded between the surfaces of the spatial domains $\Omega_{n}$ and $\Omega_{n+1}$. $P_{n}$ describes the space-time extruded boundary $\Gamma(t)$. The discretization is done with P1P1 finite elements. Figure \ref{fig:spaceslab} shows a space-time slab for a discretized 2D spatial domain.
	
	The trial and weight function spaces for each time slab of the velocity $\Big( S_{\mathbf{u}}^h \Big)_{n}$, $\Big( {V}_{\mathbf{u}}^h\Big)_{n}$ and pressure $\Big( S_p^h\Big)_{n}$, $\Big( V_p^h\Big)_{n}$ can be defined such that:
	\begin{align} 
	\Big( {S}_{\mathbf{u}}^h \Big)_{n}&= \{ \mathbf{u}^h | \, \mathbf{u}^h  \in H^{1h}\left( Q_{n} \right) | \mathbf{u}^h = \mathbf{g} \text{ on } \left( P_n\right)_{\mathbf{g}} \} \, , \\
	\Big( {V}_{\mathbf{u}}^h \Big)_{n}&= \{ \mathbf{w}^h | \, \mathbf{w}^h  \in H^{1h}\left( Q_{n} \right) | \mathbf{w}^h = \mathbf{0} \text{ on } \left( P_n\right)_{\mathbf{g}} \}  \, ,\\
	\Big(S_p^h\Big)_{n}&= \Big(V_p^h\Big)_{n}= \{ p^h | \, p^h  \in H^{1h}\left( Q_{n} \right) \} \, .
	\end{align}
	The stabilized weak formulation of the Navier-Stokes equation can be written as: Given $\left(\mathbf{u}^{h}\right)^{-}_{n}$ find $\mathbf{u}^h \in {S}_{\mathbf{u}}^h$ and $p^h \in {V}_{p}^h$ such that $\forall \mathbf{w}^h \in {V}_{\mathbf{u}}^h$ and $\forall q^h \in {V}_p^h$:
	
	\begin{align}\label{eq:spacetime-weak}
	&\left(\mathbf{w}^{h},\rho\frac{\partial \mathbf{u}^{h}}{\partial t}\right)_{Q_{n}}
	+ \left( \mathbf{w}^{h},\rho   \mathbf{u} \cdot\nabla \mathbf{u}^{h}\right)_{Q_{n}}  
	+\left(\boldsymbol{\varepsilon}\left(\mathbf{w}^{h}\right),2\eta\boldsymbol{\varepsilon}\left(\mathbf{u}^{h}\right)\right)_{Q_{n}}\nonumber \\
	&-\left(\nabla \cdot \mathbf{w}^{h}, p^{h}\right)_{Q_{n}} 
	+\left(q^{h},\nabla \cdot \mathbf{u}^{h}\right)_{Q_{n}}\nonumber \\
	&+\begin{rcases}
	\left(\left(\mathbf{w}^{h}\right)^{+}_{n},\rho\left(\left(\mathbf{u}^{h}\right)^{+}_{n}-\left(\mathbf{u}^{h}\right)^{-}_{n}\right)\right)_{\Omega_{n}}\end{rcases}\text{jump term} \nonumber \\
	&+\sum_{e=1}^{\left(n_{el}\right)_{n}} \left(\left(\rho\left(\frac{\partial\mathbf{w}^{h}}{\partial t}+\mathbf{u^{h}\cdot\nabla\mathbf{w}^{h}}\right)-\nabla\cdot\boldsymbol{\sigma}\left(\mathbf{w}^{h},q^{h}\right)\right),\tau_{M}\mathbf{r}_{M}/\rho\right)_{Q_{n}^{e}} \nonumber\\
	&+\sum_{e=1}^{\left(n_{el}\right)_{n}} \left(\nabla\cdot\mathbf{w}^{h},-\rho \tau_{C} r_C  \right)_{Q_{n}^{e}}\nonumber\\
	&=\left(\mathbf{w}^{h},\mathbf{h}^{h}\right)_{\left(P_{n}\right)_{\mathbf{h}}}+\left(\mathbf{w}^{h},\mathbf{f}\right) _{Q_{n}}\, .
	\end{align}
	Here, the notation $\left(\mathbf{u}^{h}\right)^{\pm}_{n}=\lim\limits_{\epsilon \rightarrow 0}\mathbf{u}^{h}\left(t_{n}\pm\epsilon\right)$, $\left(\mathbf{u}^{h}\right)^{-}_{0}=\mathbf{u}_{0}$, $\left(\cdot,\cdot\right)_{Q_{n}}=\int_{I_{n}} \int_{\Omega^{h}}\left(\cdot,\cdot\right)\diff\Omega \diff t$ and $\left(\cdot,\cdot\right)_{P_{n}}=\int_{I_{n}} \int_{\Gamma_{\mathbf{h}}}\left(\cdot,\cdot\right)\diff\Gamma \diff t$ is used. A jump term in Eq.~\eqref{eq:spacetime-weak} weakly enforces the continuity of the temporal solution. The stabilization terms are based on the Galerkin/least-squares (GLS) method, where $\tau_{M}$ and $\tau_{C}$ are stabilization parameters.
	Further details on the finite element method and stabilization terms, including definition of residuals $\mathbf{r}_{M}$ and $r_C$, can be found in the study of Pauli~\cite{Pauli2016}.
	
	\subsection{Finite element formulation of the level set equation} \label{sec:FEMlevelset}
	In the space-time framework, a level set equation is applied to capture the interface that is introduced in Eq.~\eqref{eq:levelset}. Let the trial and function spaces of the level set marker be defined as $\left(S^{h}_{\phi}\right)_{n}$ and $\left(V_{\Psi}^{h}\right)_{n}$, such that:
	\begin{align}
	\left(S^{h}_{\phi}\right)_{n} &= \{ \phi^h | \, \phi^h  \in H^{1h} | \phi^h = \hat{\phi}  \text{ on } \left( P_n\right)_ {\hat{\phi}} \} \, , \\
	\left(V_{\Psi}^{h}\right)_{n}&= \{ \Psi^{h} | \, \Psi^{h}  \in H^{1h} | \Psi^{h}= 0 \text{ on } \left( P_n\right)_ {\hat{\phi}} \} \,.
	\end{align}
	Then, the stabilized formulation of the level set equation can be stated as: Find $\phi^h \in \left(S^{h}_{\phi}\right)_{n}$, such that $\forall\Psi^h\in \left(V_{\Psi}^{h}\right)_{n}$
	\begin{align}
	&\left(\Psi^{h}, \frac{\partial \phi^{h}}{\partial t} + \mathbf{u}^{h}\cdot \nabla \phi^{h} \right)_{Q_{n}} +
	\left(\left(\Psi^{h}\right)^{+}_{n}\left(\left(\phi^{h}\right)^{+}_{n}-\left(\phi^{h}\right)^{-}_{n}\right)\right)_{\Omega_{n}} \nonumber\\
	&+ \sum_{e=1}^{\left(n_{el}\right)_{n}}\left(\left(\frac{\partial \Psi^{h}}{\partial t}+\mathbf{u}^{h}\cdot \nabla \Psi ^{h}\right)\tau_{L}\left(\frac{\partial \phi^{h}}{\partial t} + \mathbf{u}^{h}\cdot \nabla \phi^{h}\right)\right)_{Q_{n}^{e}}=0 \, ,
	\label{eq:levelset_weak}
	\end{align}
	where $\tau_{L}$ is the stabilization parameter for the level set equation. For more details refer to Sauerland and Fries~\cite{Sauerland2013a}
	\subsection{Space-Time elements}
	Most often, space-time meshes are extruded in the temporal direction from a spatial mesh, forming prisms. In a further step the prisms can be transformed to simplex elements. In this paper, simulations are done with both prismatic space-time, also called flat space time (FST), meshes and simplex space-time (SST) meshes. Below, it is explained how space-time meshes are generated and refined, following the developments in the study of Behr \cite{Behr2008} closely.
	\subsubsection{Flat space-time (FST)}
\begin{figure}[h!]
		\centerline{
			\begin{subfigure}[c]{0.2\textwidth}
				\centerline{
					\includegraphics[width=\textwidth]{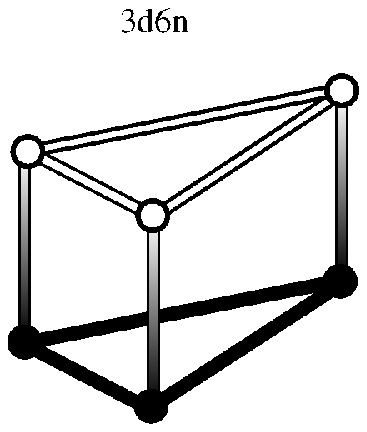}}
				\caption{3D prism element extruded from a 2D triangular element.}
				\label{fig:prismaticelementA}	
			\end{subfigure}
			\hspace{1.0cm}
			\begin{subfigure}[c]{0.065\textwidth}
				\centerline{
					\includegraphics[width=\textwidth]{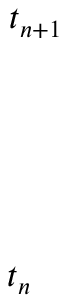}}
				\caption*{}
			\end{subfigure}
			\hspace{1.0cm}
			\begin{subfigure}[c]{0.2\textwidth}
				\centerline{
					\includegraphics[width=\textwidth]{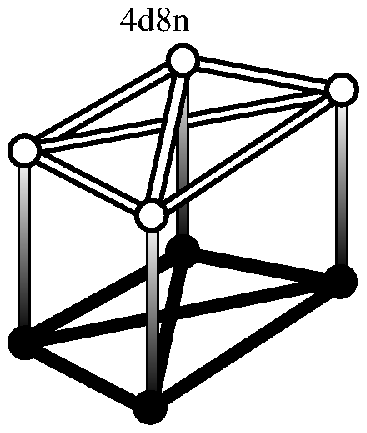}}
				\caption{4D space-time element extruded from a 3D tetrahedron.}
				\label{fig:prismaticelementB}	
		\end{subfigure}}
		\caption{Prismatic space-time elements.}
		\label{fig:prismaticelement}
	\end{figure}

	An unstructured mesh in space but structured in time is called a semi-unstructured space-time approach. This approach does not allow any specific temporal refinement for certain areas of interest, e.g., the time nodes are uniformly distributed throughout the domain. Thus, the space-time slab is equivalent to the time step of a semi-discrete method. The mesh consists of prismatic elements, which fill the time slab between the time levels $t_{n}$ and $t_{n+1}$. Figure~\ref{fig:prismaticelementA} illustrates a 2D triangular element that is extruded to a 3D space-time 6-noded prism, referred to as 3d6n. For the 3D case, a tetrahedron is extruded to create a 4D space-time mesh with an 8-noded space-time element, called 4d8n element, that can be seen in Fig.~\ref{fig:prismaticelementB}. 
	\subsubsection{Simplex space-time (SST)}
\begin{figure}[h!]
		
		\centerline{
			\captionsetup[subfigure]{font=normalsize,labelfont=normalsize}
			\begin{subfigure}[c]{0.15\textwidth}
				\centerline{
					\includegraphics[width=\textwidth]{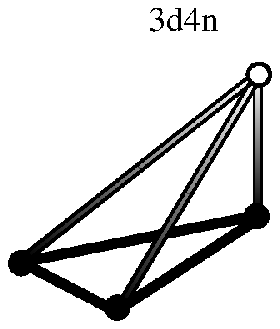}}
				\caption{3D simplex element.}
				\label{fig:simplexelementA}	
			\end{subfigure}
		\hspace{1.0cm}
			\begin{subfigure}[c]{0.065\textwidth}
				\centerline{
					\includegraphics[width=\textwidth]{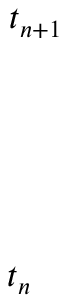}}
				\caption*{}
			\end{subfigure}
			\hspace{0.5cm}
			\begin{subfigure}[c]{0.15\textwidth}
				\centerline{
					\includegraphics[width=\textwidth]{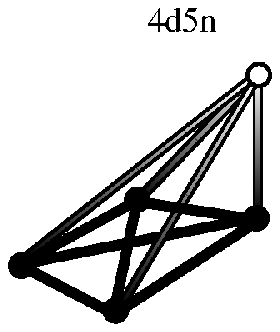}}
				\caption{4D pentatope element.}
				\label{fig:simplexelementB}	
		\end{subfigure}}
		\caption{Simplex space-time elements.}
		\label{fig:simplexelement}
	\end{figure}

	Space-time formulations can also be based on fully unstructured meshes (unstructured not only in space but also in time) to efficiently resolve problems with complex boundaries, like the interface in a two-phase flow. The extruded prism is split into simplices to obtain an unstructured mesh in time. This allows an unstructured refinement in time. When the prism-type elements are subdivided, the mesh consists of simplex-type elements; in the 2D case, this leads to a 3d4n, similar to a tetrahedron, which can be seen in Fig.~\ref{fig:simplexelementA}. A 3D spatial mesh renders a 4d5n element, a so-called pentatope (compare Fig.~\ref{fig:simplexelementB}). Each prism is independently divided into simplices according to the Delaunay algorithm. This way, a finer temporal resolution can be applied in certain parts of the domain. 
	
\begin{figure}[h!]
		\centerline{
			\includegraphics[width=0.2\textwidth]{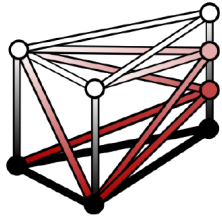}}
		\caption{Simplexes splitting up a prism element for temporal refinement.}
		\label{fig:simplexrefinement}		
	\end{figure}

	By dividing a prism element into multiple simplexes, one side of the element can be more refined than the other side. Figure~\ref{fig:simplexrefinement} illustrates such a refinement, where one side has three subdivisions and the other two sides have only one interval. This way it is possible to have unstructured refinement in the temporal direction. 
	
\begin{figure}[h!]
		\centerline{
			\includegraphics[width=0.2\textwidth]{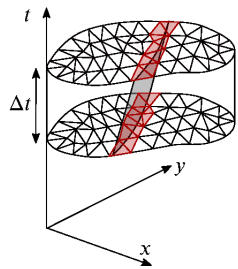}}
		\caption{Elements close to the interface of a two-phase flow that will be temporally refined.}
		\label{fig:spacetimeinterface}		
	\end{figure}

	The temporal refinement is interesting for some areas, such as the interface in a two-phase flow. Areas that are close to the interface can be resolved with smaller elements. Figure~\ref{fig:spacetimeinterface} shows an interface of a two-phase flow. The elements close to the interface are marked to be refined in temporal direction \cite{Karyofylli2018}.

	\subsection{No-slip boundary condition for two-phase flow} \label{sec:no-slip}
	The no-slip boundary condition on the wall of a two-phase flow can lead to difficulties. As the position of the interface of the two fluids is calculated by the level set method, the resulting point of the transition between the light and the heavy fluid depends on the velocity. Due to the no-slip condition, the velocity at the wall is zero, therefore, preventing the values of the level set marker to alternate. This leads to a phenomenon where the nodes stay with a marker value of the light fluid, whereas the upper nodes already change their values to the heavy fluid. 
	To avoid this, we use Navier slip condition for the bottom boundary condition. The Navier slip is a mixture of Dirichlet and Neumann boundary condition. This way, instead of setting the boundary either entirely as no-slip or slip condition, a slip condition can be defined close to the interface. The Navier slip boundary condition can be stated as:
	\begin{align}
	\mathbf{n}\cdot \mathbf{u} = 0 \quad \text{on} \quad \boldsymbol{\Gamma}_{slip}\,, \\
	\mathbf{t}\cdot \sigma\left(\mathbf{u},p\right)\cdot\mathbf{n}= \beta \mathbf{t}\cdot\mathbf{u} \quad \text{on}\quad \boldsymbol{\Gamma}_{slip}\,,
	\end{align}
	here $\mathbf{n}$ is the normal and $\mathbf{t}$ holds the tangent vectors at the boundary. The coefficient $\beta$ can be seen as a penalty parameter that is small close to the interface and has high values everywhere else. 
	\section{Results}
	In the following, two different experiment settings for dense granular flow are presented. First, a column collapse is analyzed to show the differences in the performance of FST and SST. Hereby, a low and a high column aspect ratio are considered.  Afterwards, we compute a more complex test case. A dam break down an inclined plane is simulated and verified by the impact time and force on a rigid obstacle. The simulation uses SST with an unstructured mesh in time around the interface.
	
	\subsection{Column collapse} \label{sec:ColumnCollapse}
	The 2D column collapse benchmark consists of a column of a heavy, dense granular fluid surrounded by a light, air-like fluid. The viscosity of the heavy fluid is modeled as a non-Newtonian fluid applying the $\mu$($I$)-rheology, whereas the light fluid is assumed to have Newtonian fluid properties. 
	
\begin{figure}[h!]
		\centering
		\begin{subfigure}[c]{0.45\textwidth}
			\centerline{
				\includegraphics[width=\textwidth]{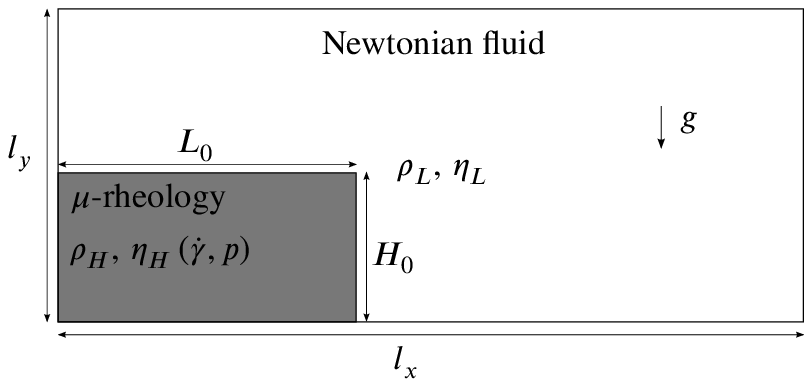}}
			\normalsize{ \caption{$a=0.5$}}
		\end{subfigure}
		\hspace{0.5cm}
		\begin{subfigure}[c]{0.45\textwidth}
			\centerline{
				\includegraphics[width=\textwidth]{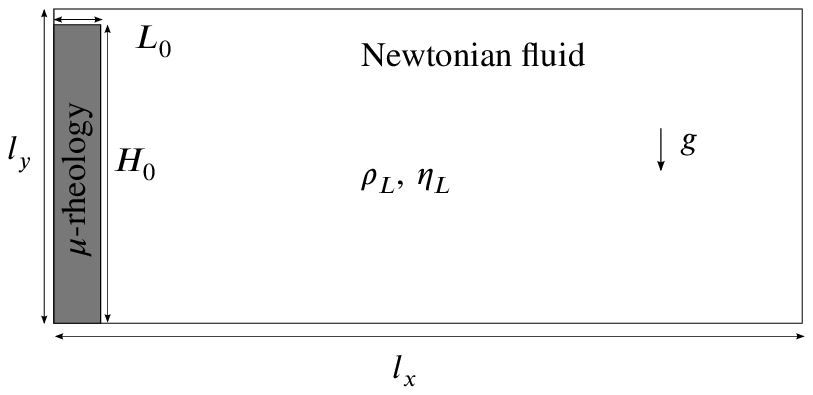}}
			\caption{$a=6.26$}
		\end{subfigure}
		\caption{Dimensions and configuration of the column collapse.}
		\label{fig:geo2D}
	\end{figure}

	We simulate two different column aspect ratios that can be seen in Fig.~\ref{fig:geo2D}. The dimensions of the experiment are $l_{x}=5.5$, $l_{y}=1.1$. The first column has a relatively low aspect ratio of $a=H_{0}/L_{0}=0.5$.  The second geometry consists of a thin column with a higher aspect ratio of $a=H_{0}/L_{0}=6.26$. The interface is calculated by the level set method from Sec.~\ref{sec:FEMlevelset}. The heavy fluid has a negative level set marker, whereas the light fluid has a positive marker value. The interface is defined at the level set marker $\phi=0$. The side and top boundaries have slip conditions. On the bottom boundary, Navier slip is applied.
	
	The rheological parameters of the $\mu$($I$)-rheology are $\mu_{s}=0.32$, $\mu_{2}=0.60$, and $I_{0}=0.4$. The normalized grain diameter is assumed to be ${d}=H_{0}/\sqrt{a \cdot N_{g}}$, where $N_{g}$ is the number of grains in the column. The normalization is taken from Lagr\'{e}e \textit{et al.} \cite{Lagree2011}. The first column has $N_{g}^{0.5}=3407 $ grains, and the second column has $N_{g}^{6.26}=6036$ grains. All results are normalized with the characteristic time $t^{*}=\sqrt{H_{0}/g}$, the characteristic length $L^{*}=L_{0}$ and the characteristic density $\rho^{*}=\rho$. 
	The density of the light fluid is normalized by the grain density and is ${\rho_{L}}=10^{-3}$. In the following example, the density of the heavy fluid and the grain density are $\rho_H=\rho_g=1.0$, and a normalized gravity $g=1.0$ is used. The normalized viscosity of the light fluid is ${\eta_{L}}=10^{-4}g^{1/2}L_{0}^{3/2}\rho_{H}$. The partial regularization of Barker and Gray \cite{Barker2017}, which is given in Eq.~\eqref{eq:barker2}, is applied. Besides this regularization, viscosity thresholds are added to both formulations, as described in Eq.~\eqref{eq:viscositylimit}. The maximum viscosity limit is $\eta_{max}=1000.0$. The minimum viscosity of the heavy fluid needs to be higher than the light fluid's viscosity to avoid mixing. The minimum viscosity is assumed as a multiple of the viscosity of the light fluid. Depending on the geometry, a minimum viscosity of $\eta_{min}^{0.5}=20 \;\eta_{L}$,  and $\eta_{min}^{6.26}=200 \;\eta_{L}$ is used.
	
	The spatial mesh consists of triangular elements. The spatial mesh of the column with an aspect ratio of $a=0.5$ has $15,456$ nodes and is then extruded to a 3D space-time mesh with $30,250$ elements and $30,912$ nodes. For the high aspect ratio column with $a=6.26$, the spatial mesh consists of $243,321$ nodes and is extruded to a 3D space-time mesh with $486,642$ nodes and $484,000$ elements.

	For both geometries, three different temporal discretizations are compared: Firstly, an FST discretization with $250$ time slabs, where $\Delta t = 0.02$, denoted as FST $\Delta t = 0.02$. Secondly, an FST consisting of 50 time slabs, where $\Delta t =0.1$ (FST $\Delta t = 0.1$). Thirdly, an unstructured SST discretization that temporally refines the mesh at the interface. $50$ time slabs are simulated and the time slab size varies between $0.02-0.1$ with 1 to 5 elements in time.  The final time of the simulations are $\bar{t}_{final}=5$
	
\begin{figure}[h!]
		\centering
		\begin{subfigure}[c]{0.9\textwidth}
			\centerline{
				\includegraphics[width=\textwidth]{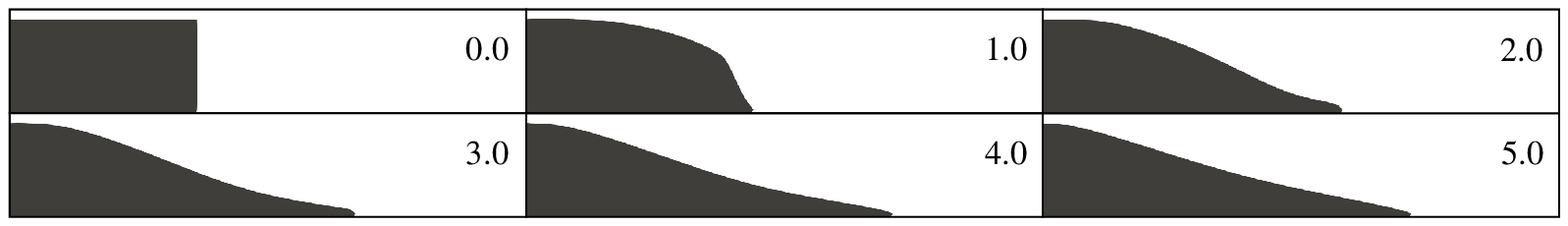}}
			\caption{$a=0.5$}
			\label{fig:ColumnSnapshots1}
		\end{subfigure}
		\hspace{0.5\textwidth}
		\hspace{0.5cm}
		\begin{subfigure}[c]{0.9\textwidth}
			\centerline{
				\includegraphics[width=\textwidth]{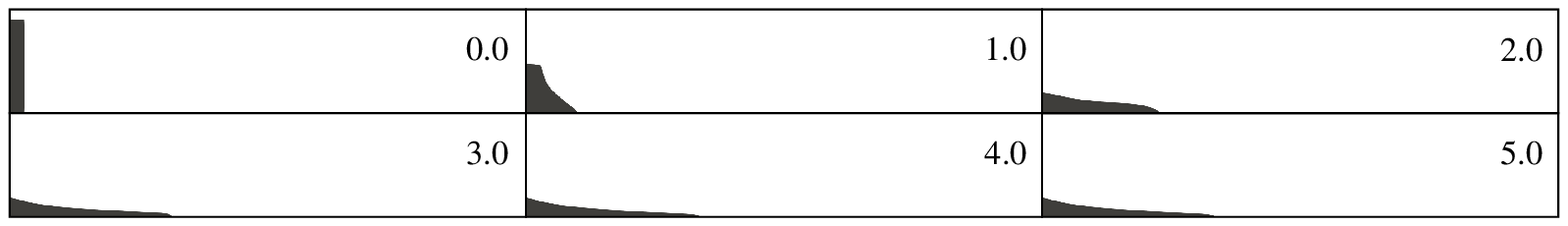}}
			\caption{$a=6.26$}
			\label{fig:ColumnSnapshots2}
		\end{subfigure}
		\caption{Snapshots of the column collapse.}
		\label{fig:ColumnSnapshots}
	\end{figure}

	In Fig.~\ref{fig:ColumnSnapshots}, flow front snapshots of the column collapse are presented. As the flow fronts of all time discretizations are similar, we only show the flow fronts of the FST $\Delta t = 0.02$ time discretization. Figure~\ref{fig:ColumnSnapshots1} shows the column with an aspect ratio of $a=0.5$, and Fig.~\ref{fig:ColumnSnapshots2} the column with $a=6.26$   at the times $\bar{t}=\{0.0, \, 1.0, \, 2.0, \, 3.0, \, 4.0, \,  5.0\}$. Over time the column collapses and comes to rest towards the end of the simulation.
	
\begin{figure}
		\begin{subfigure}[c]{\textwidth}
			\centerline{\includegraphics[width=0.6\textwidth]{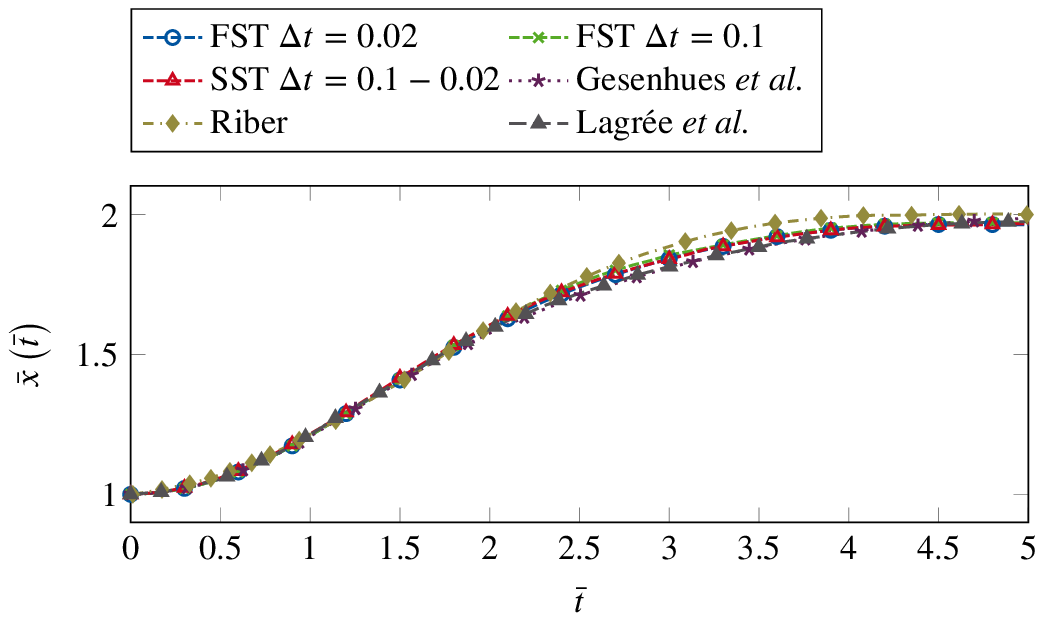}}
			\caption{$a=0.5$}
			\label{fig:ColumnCollapseVerification1}
		\end{subfigure}
		\begin{subfigure}[c]{\textwidth}
			\centerline{
				\includegraphics[width=0.6\textwidth]{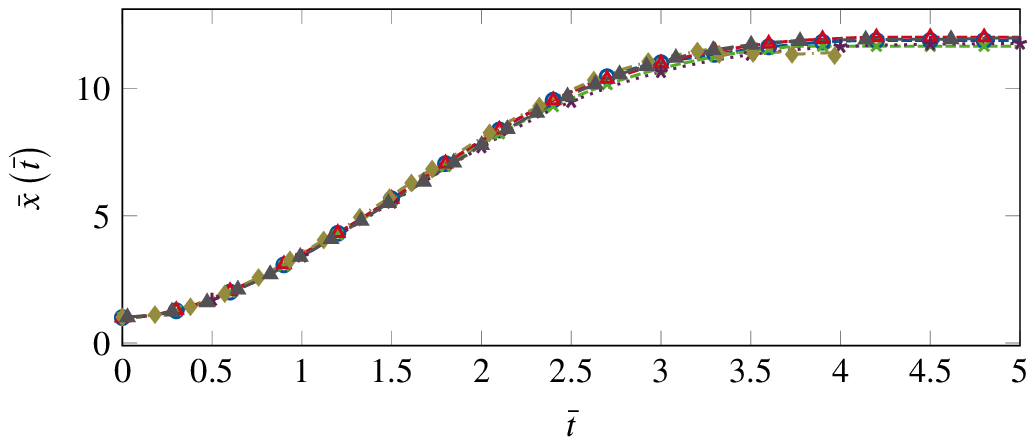}}
			\caption{$a=6.26$}
			\label{fig:ColumnCollapseVerification2}
		\end{subfigure}
		\caption{Flow front position $\bar{x}=x/L_{0}$ of the heavy fluid at $\bar{t}=t/\sqrt{H_{0}/g}$.}
		\label{fig:ColumnCollapseVerification}
	\end{figure}

	Figure~\ref{fig:ColumnCollapseVerification1} shows the normalized flow front positions $\bar{x}=x/L_{0}$ over the normalized time $\bar{t}=t/\sqrt{H_{0}/g}$ of the column collapse. The flow front position is the maximum $x$-coordinate of the heavy fluid that is evaluated for each time step. 
	The three different space-time discretizations are compared and verified with the results of other authors who use semi-discrete time discretizations. Overall, all results based on a space-time formulation agree with the results presented by Lagr\'{e}e \textit{et al.}~\cite{Lagree2011}, Riber~\cite{Riber2017}, and  Gesenhues \textit{et al.}~\cite{Gesenhues2018}. Small differences might occur due to the higher diffusivity of the space-time approach.

\begin{figure}[h!]
		\centering
		\begin{subfigure}[c]{0.8\textwidth}
			\centerline{
			\includegraphics[width=\textwidth]{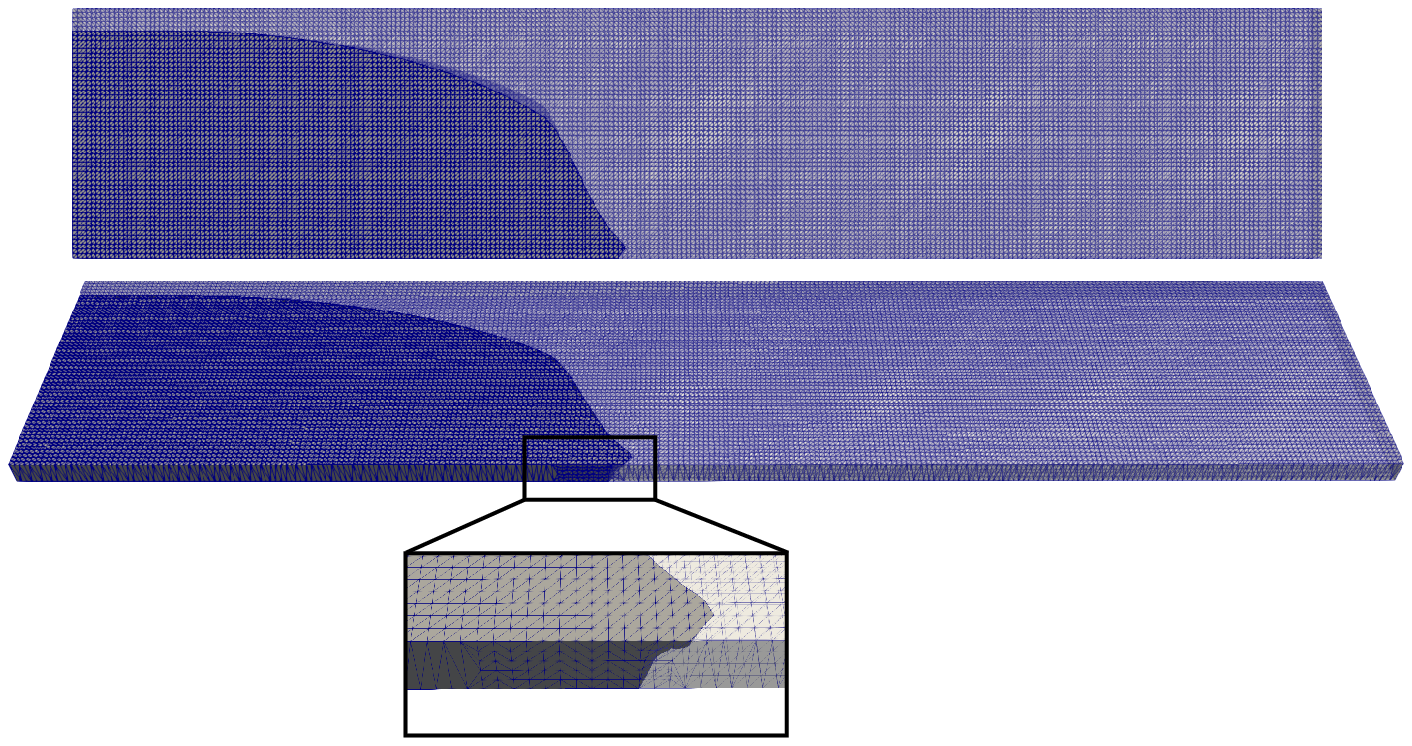}}
			\caption{$a=0.5$}
			\label{fig:ColumnSTmesh1}
		\end{subfigure}
		\hspace{0.5\textwidth}
		\hspace{0.5cm}
		\begin{subfigure}[c]{0.8\textwidth}
			\centerline{
			\includegraphics[width=\textwidth]{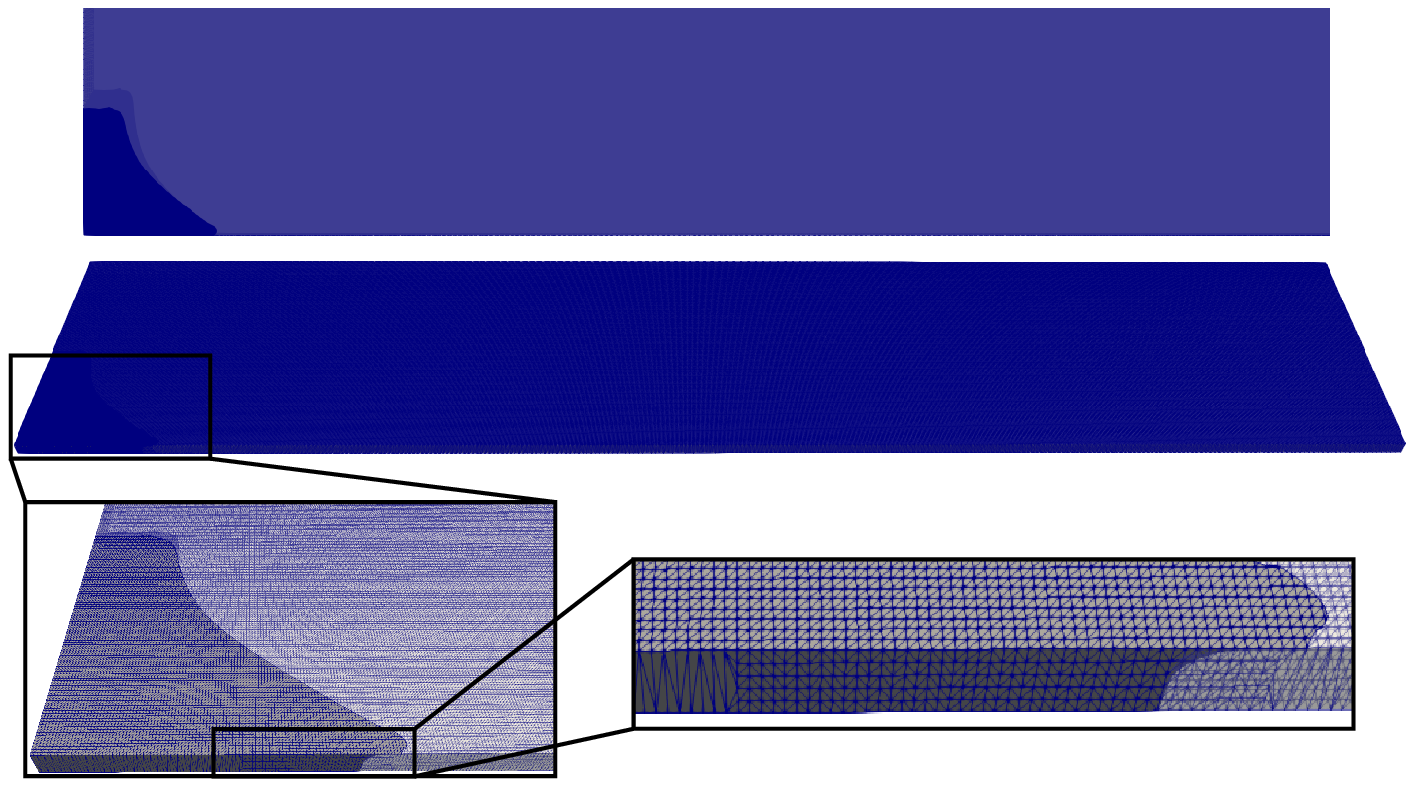}}
			\caption{$a=6.26$}
			\label{fig:ColumnSTmesh2}
		\end{subfigure}
		\caption{SST mesh at $\bar{t}=1.0$ column collapse.}
		\label{fig:ColumnSTmesh}
	\end{figure}

	In Fig.~\ref{fig:ColumnSTmesh}, a section of the SST mesh at $\bar{t}=1.0$ is illustrated.  For better visualization, we show a zoomed section from the side close to the interface. From the 2D spatial mesh, a 3D space-time mesh is generated.  It can be seen that close to the interface, the mesh has five layers of refinement in the temporal direction.  At $\bar{t}=1.0$, the mesh of the column with $a=0.5$ includes $108,470$ elements and $33,900$ nodes, whereas the mesh of  the column with $a=6.26$ contains  $1,558,656$ elements and $504,578$ nodes. Note that the two geometries use different spatial meshes. For the SST meshes, the number of elements and nodes depends not only on the spatial mesh but also on the interface length (or extent).
	
	Table~\ref{tab:ColumnPerformance1} displays the number of time steps, nodes per step, elements per step, and the performance times for the system formation and system solution for the three discretizations of the column collapse simulation with aspect ratio $a=0.5$. The FST $\Delta t = 0.02$ has 250 time steps, whereas the other discretizations only have 50 time steps. However, the SST discretization has a higher number of nodes and elements. It has five times as many levels of refinement for the interface region. Therefore, the number of nodes and elements depends on the interface area. Because of the high number of time steps, the FST $\Delta t = 0.02$ has the highest system formation and solution time. The FST  $\Delta t = 0.1$  has a comparable performance to the SST simulation. However, the unstructured SST resolves the area of the interface with finer elements. Even though the SST discretization has significantly more elements, it has the lowest system solution time. Due to the mesh refinement close to the interface, the simulation converges faster.
	
	\begin{center}
		\begin{table}[t]%
			\centering
			\caption{Number of time steps, nodes, elements and performance times for the column aspect ratio $a=0.5$.\label{tab:ColumnPerformance1}}%
			\begin{tabular*}{500pt}{@{\extracolsep\fill}lccccc@{\extracolsep\fill}}
				\toprule
				& \textbf{Time steps}  & \textbf{Nodes per step} & \textbf{Elements per step}  & \textbf{System formation, \unit{s}}  & \textbf{System solution, \unit{s}} \\
				\midrule
				FST $\Delta t = 0.02$ & 250 & 30,912  & 30,250 & 499.44  & 1,526.62  \\
				FST $\Delta t = 0.1$ &  50 & 30,912  & 30,250  & 245.19  & 746.6   \\
				SST $\Delta t = 0.1-0.02$ & 50  & $\sim$34,185 & $\sim$110,219  & 260.88 & 667.68    \\
				\bottomrule
			\end{tabular*}
		\end{table}
	\end{center}

	Table~\ref{tab:ColumnPerformance1} states the number of time steps, nodes per step, elements per step, and the performance times for the system formation and system solution for the three discretizations for the column with aspect ratio $a=6.26$. Again, the FST $\Delta t = 0.02$ has the highest system formation and system solution times.  FST $\Delta t = 0.02$  has 250 time steps, whereas the other two discretizations only have 50 time steps. Even though the SST has averagely more nodes and elements per time step is has the best performance regarding the computation time.  Temporal mesh refinement improves the convergence of the system and therefore leads to a faster system solution.
	\begin{center}
		\begin{table}[t]%
			\centering
			\caption{Number of time steps, nodes, elements and performance times for the column aspect ratio $a=6.26$.\label{tab:ColumnPerformance2}}%
			\begin{tabular*}{500pt}{@{\extracolsep\fill}lccccc@{\extracolsep\fill}}
				\toprule
				& \textbf{Time steps}  & \textbf{Nodes per step} & \textbf{Elements per step}  & \textbf{System formation, \unit{s}}  & \textbf{System solution, \unit{s}} \\
				\midrule
				FST $\Delta t = 0.02$ & 250 & 486,642  & 484,000  & 3,321.58  & 11,913.24 \\
				FST $\Delta t = 0.1$ &  50 & 486,642  & 484,000  & 2,044.12  & 6,714.08  \\
				SST $\Delta t = 0.1-0.02$ & 50  & $\sim$509,721  & $\sim$1,589,276  & 1,917.94 & 5,895.41    \\
				\bottomrule
			\end{tabular*}
		\end{table}
	\end{center}
	
	\subsection{Inclined dam break with a rigid obstacle} \label{sec:Bulldoze}
	The second test case is a dam break that flows down an inclined plane and impacting a rigid obstacle. The high velocity and viscosity gradients over time require a well refined mesh and a small time step, especially around the interface. We use the simplex space-time formulation to create an unstructured mesh in time with smaller effective time steps around the interface.
	
	In their study, Moriguchi \textit{et al.}~\cite{Moriguchi2009} perform a physical experiment with sand and measure the impact time and force on the obstacle. They also run a numerical simulation using a Bingham-Mohr-Coulomb (BMC) model. The model depends on cohesion, pressure, shear rate, and a friction angle. A Bingham-Mohr-Coulomb model is also, like the $\mu$($I$)-rheology, a friction-based viscoplastic material model. The numerical experiment determines the hydrodynamic impact force on the obstacle.
	
	The laboratory experiment consists of a sandbox located in the upper part, a flume, and an obstacle that is equipped with sensors to measure the history of the impact force.  At the time $t=0.0$, the box with uniform dry sand is opened. The sand runs down the rim, and reaches an obstacle further down, which eventually stops the flow.
	
	In our study, the same experiment is simulated; however, we apply the $\mu$($I$)-rheology. The material parameters of the $\mu$($I$)-rheology are fitted to the parameters of the Bingham-Mohr-Coulomb model ~\cite{Gesenhues2020}.
	
\begin{figure}[h!]
		\centerline{
			\includegraphics[width=0.45\textwidth]{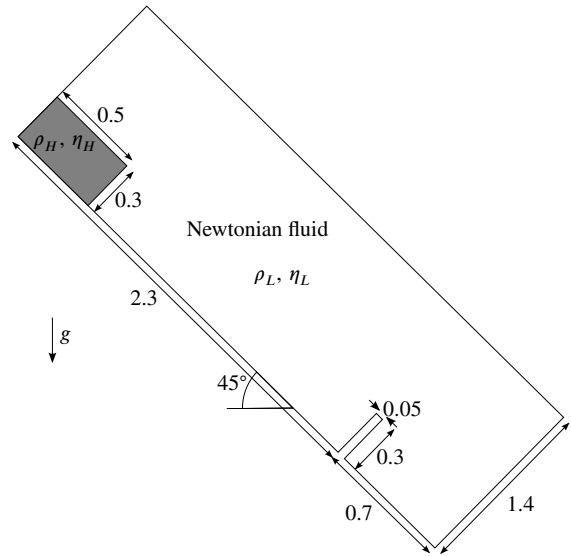}}
		\caption{Dimensions and configuration of the dam break down an inclined plane.}
		\label{fig:BulldozeGeo}
	\end{figure}

	Figure~\ref{fig:BulldozeGeo} shows the configuration of the experiment, analogous to the set up of Moriguchi \textit{et al.}~\cite{Moriguchi2009}. The inclination angle is $45^{\circ}$. The box of sand has dimensions of $\unit[0.5\times 0.3]{m}$. The obstacle is located at a distance of $\unit[2.3]{m}$ from the wall, has a height of $\unit[0.3]{m}$ and a thickness of $\unit[0.05]{m}$. The dimensions of the outer box are $\unit[3.0\times1.4]{m}$. The density of the light fluid is $\rho_{L}=\unit[1.25]{kg/m^{3}}$ and the sand has a density of $\rho_{H}=\unit[1,379]{kg/m^{3}}$. The viscosity of the light fluid is $\eta_{L}=\unit[1.8\times10^{-5}]{Pa\,s}$. The material parameters of the sand are $\mu_{s} = 0.7$, $\mu_{2}= 0.93$, $I_{0}=0.6$ and a grain diameter $d=0.0009$. The left side wall and the bottom wall, as well as the obstacle walls, have a no-slip boundary condition. At the top and the right wall, slip boundary conditions are applied. The spatial mesh consists of triangular elements and has $71,073$ nodes. The resulting space-time mesh without refinement has $422,982$ elements and $142,146$ nodes. The size of the time slab is in between $\Delta t = 0.001-0.0002$ with 1 to 5 elements in time. 
	
\begin{figure}[h!]
		\begin{flushleft}
			\centerline{
				\includegraphics[width=0.7\textwidth]{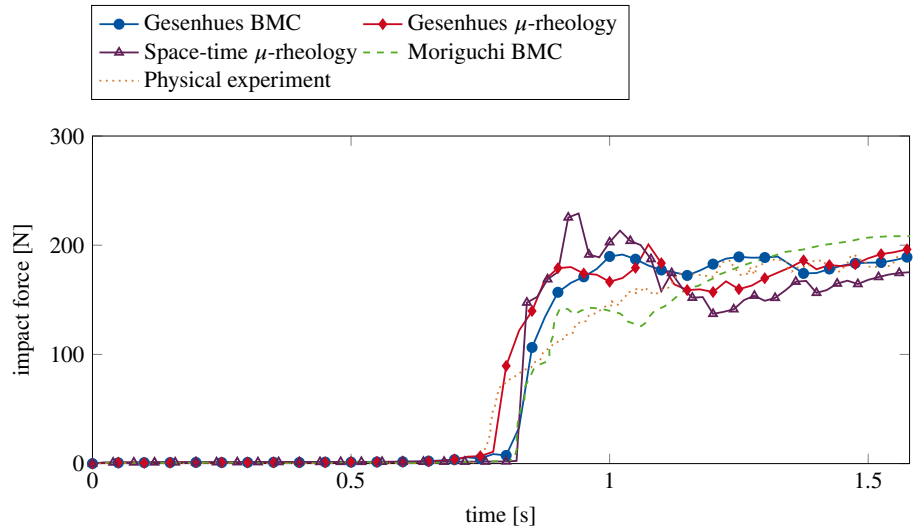}}
		\end{flushleft}
		\caption{Impact force on the rigid obstacle over time.}
		\label{fig:BulldozeVerification}
	\end{figure}

	Figure~\ref{fig:BulldozeVerification} shows the impact force on the obstacle over time. The impact force is considered proportional to the square of the flow velocity. The results of the impact force of the physical and numerical experiments using the BMC model (with a friction angle of $35^{\circ}$) of Moriguchi \textit{et al.}~\cite{Moriguchi2009} are shown, as well as the numerical simulations of the BMC and $\mu$($I$)-rheology model of Gesenhues~\cite{Gesenhues2020}.  The impact force is computed by integrating the pressure on the wall of the obstacle. The peaks occur due to the mesh refinement, which changes the element size close to the wall. Between $\unit[0.7-0.8]{s}$ the sand reaches the obstacle. The results of the simplex-space-time formulation agree well with the other simulations and the physical experiment. The overshoot after reaching the obstacle of the numerical experiments can be related to the dynamics of the impact of a model without elasticity. The peak is then depending on the dissipation and time discretization.
	
\begin{figure}[h!]
		\centerline{
			\includegraphics[width=0.9\textwidth]{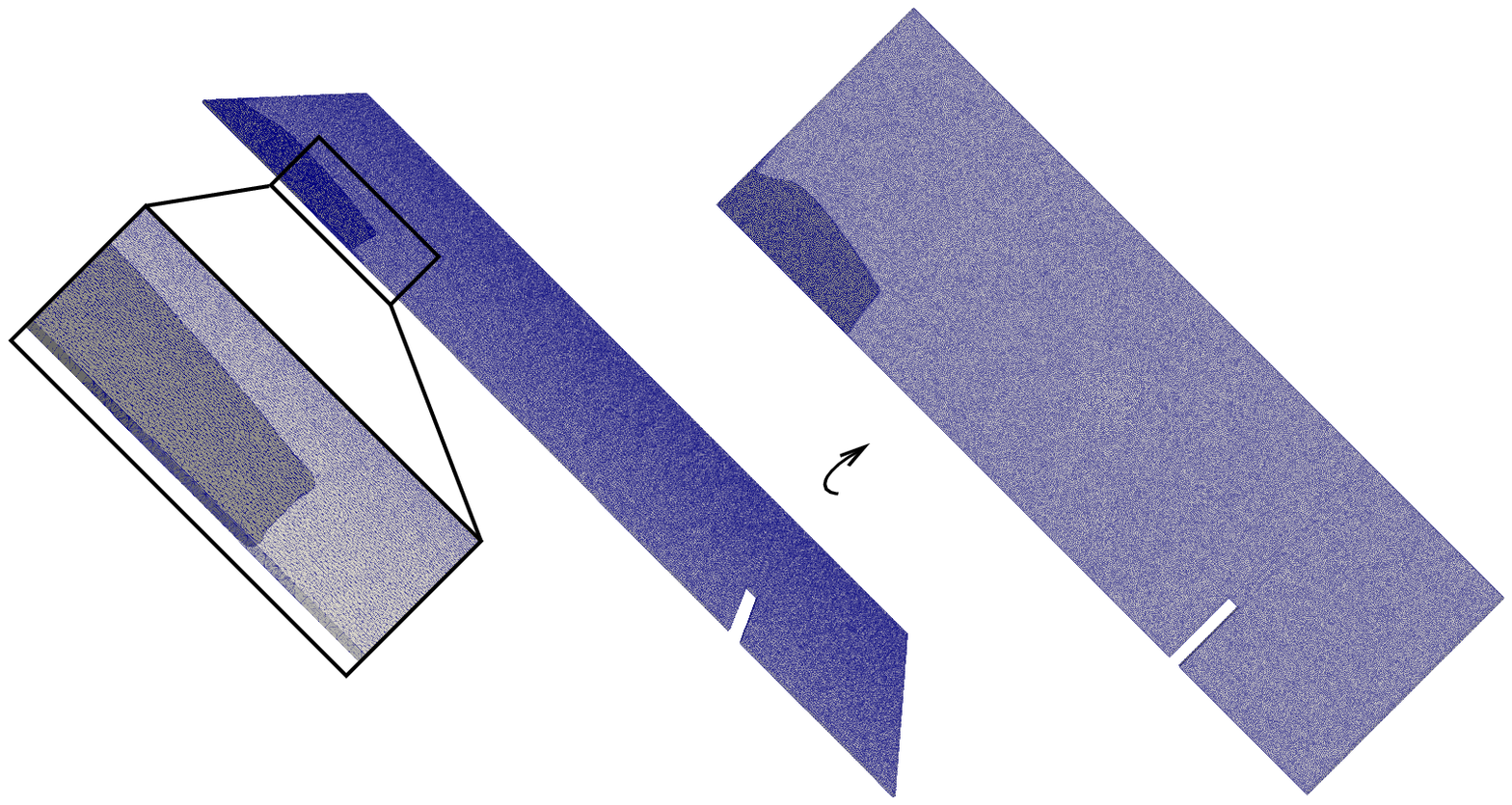}}
		\caption{Space time mesh at $t=0.2 \unit{s}$.}
		\label{fig:BulldozeMesh}
	\end{figure}

	Figure~\ref{fig:BulldozeMesh} shows the SST mesh at time $t=0.2 \unit{s}$. At this time step, the mesh has $54,428$ elements and $16,158$ nodes. Close to the interface, the mesh has five refinement levels in the temporal direction. Away from the interface, the mesh only has one element in the temporal direction.
	
\begin{figure}[h!]
		\centering
		\begin{subfigure}{0.5\textwidth}
			\centerline{
				\includegraphics[width=\textwidth]{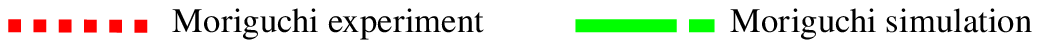}}
		\end{subfigure}
		
		\vspace{0.5cm}
		\begin{subfigure}{0.6\textwidth}
			\centerline{
				\includegraphics[width=\textwidth]{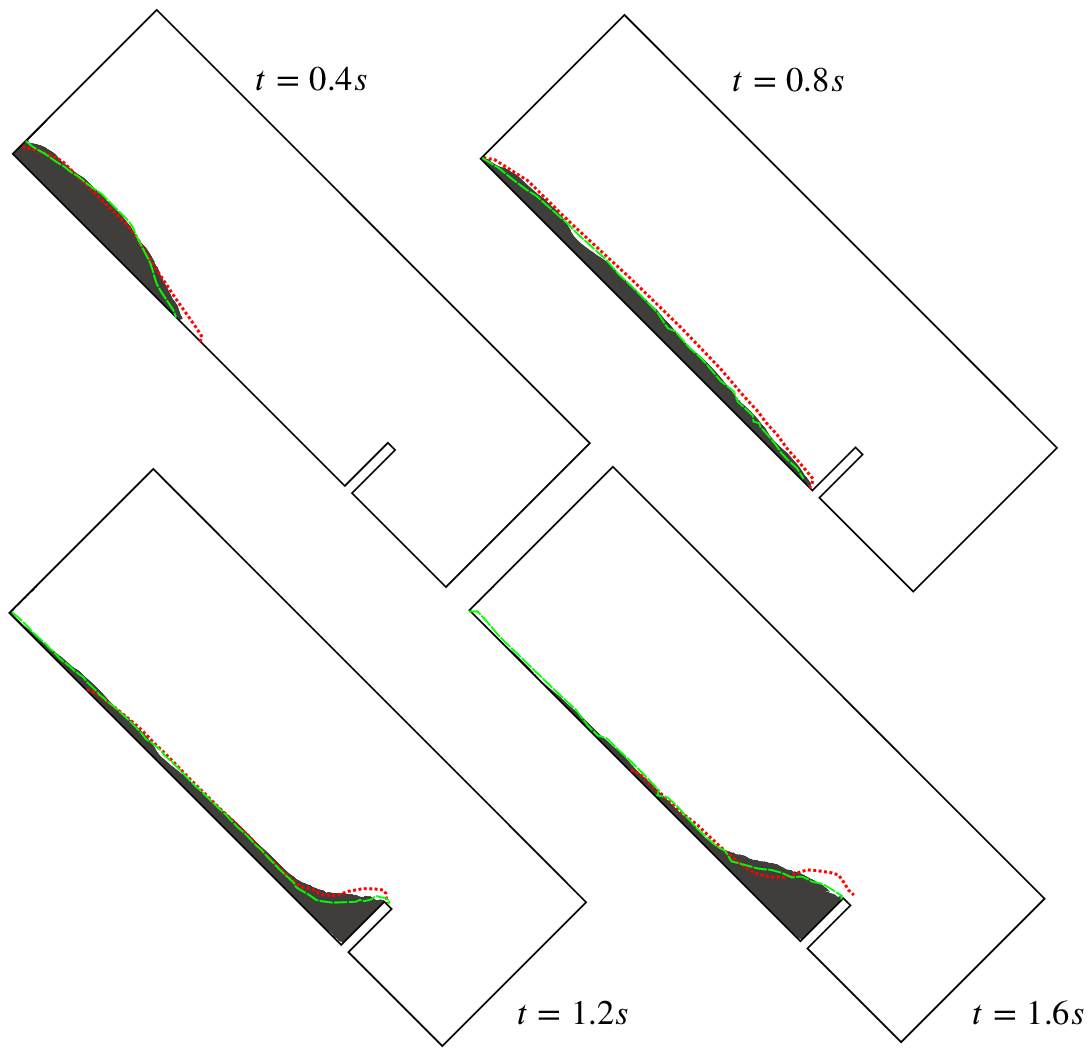}}
		\end{subfigure}
		\caption{Flow fronts of the dam break at $t=\{0.4$, $0.8$, $1.2$, $1.6\}$$ \unit{s}$.}
		\label{fig:BulldozeClip}
	\end{figure}

	In Figure~\ref{fig:BulldozeClip}, selected snapshots of the simulation results of the flow front can be seen. The dark gray area represents the heavy fluid and the white area the light fluid. The red line marks the flow front that is observed by Moriguchi \textit{et al.}~\cite{Moriguchi2009} in the physical experiment. The green line shows the flow front of their numerical simulation using a Bingham-Mohr-Coulomb model. In general, our results compare well with the results of Moriguchi \textit{et al.}~\cite{Moriguchi2009}.
	
\begin{figure}[h!]
		\centerline{
			\includegraphics[width=0.6\textwidth]{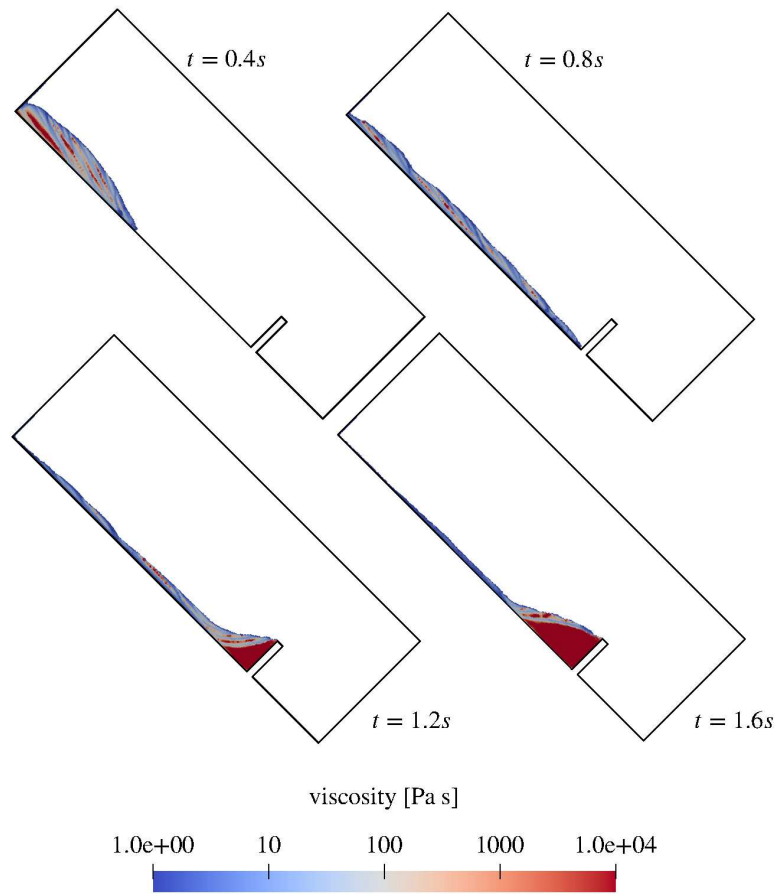}}
		\caption{Viscosity fields of the dam break at $t=\{0.4$, $0.8$, $1.2$, $1.6\}$$ \unit{s}$.}
		\label{fig:BulldozeViscosity}
	\end{figure}

	Figure~\ref{fig:BulldozeViscosity} illustrates the viscosity field of the heavy fluid at times $t=\{0.4$, $0.8$, $1.2$, $1.6\}$$ \unit{s}$. Close to the bottom, the viscosity is higher, due to lower shear rates and higher pressures. Later, when the sand is at rest at the obstacle, a high viscosity can be observed in the corner. Overall, some instabilities of the viscosity field in the form of shear bands can be seen.
	
	Table~\ref{Tab:Bulldoze} sums up the average nodes per time steps and performance times. The mesh has an average of $164,044$ nodes per time step. Due to the refinement in the temporal direction, this number is higher than the unrefined mesh. The formation of the system takes $\unit[10,492]{s}$ and the system solution $\unit[23,303]{s}$.
	
	\begin{center}
		\begin{table}[t]%
			\centering
			\caption{Number of time steps, nodes, elements and performance times for the inclined dam break.\label{Tab:Bulldoze}}%
			\begin{tabular*}{500pt}{@{\extracolsep\fill}lccccc@{\extracolsep\fill}}
				\toprule
				& \textbf{Time steps}  & \textbf{Nodes per step} & \textbf{Elements per step}  & \textbf{System formation, \unit{s}}  & \textbf{System solution, \unit{s}} \\
				\midrule
				SST & 1600  & $\sim$164,044  & $\sim$ 551,537  & 10,728.22 & 23,303.33    \\
				\bottomrule
			\end{tabular*}
		\end{table}
	\end{center}

	\section{Conclusion}\label{sec:Conclusion}
	This study uses a space-time framework to simulate a two-phase flow of dense granular material and a light Newtonian fluid. We perform two different numerical experiments: a column collapse with a low and high aspect ratio and a dam break down an inclined plane.
	The interface is refined in the temporal direction with simplex elements that allow unstructured meshes in time. This results in a very well temporal refinement around the interface and, therefore, good numerical results. The results are verified with existing studies. 
	For the column collapse, we use FST, as well as SST. For all simulations, the flow fronts agree well with the results of other authors. SST needs less time for the system solution of the simulation than FST discretization with a similar time slab size in interest regions. The inclined dam break using the SST framework shows good agreement with the physical experiment and numerical simulations using semi-discrete frameworks. Furthermore, the SST mesh allows us to simulate the dam break with a well refined interface region in time combined with a reasonable computation time.
	
	In this study, we only used unstructured temporal refinement for the area of the interface. The level set marker is used as an indicator of whether the mesh should be refined or not. Refinement close to the interface makes sense due to the discontinuities occurring here. However, temporal refinement could also be used in other regions of interest, such as areas with high viscosity or velocity gradients. These areas would need to be determined by error estimators.

	\bibliography{paperOct}
	
\end{document}